\newif\ifjournal
\journaltrue

\ifjournal
  \documentclass{aa}
  \usepackage{graphicx,times}
\else
  \documentclass{paper}
\fi

\begin{document}


\title{Correlation of the magnetic field and the intra-cluster gas density in
galaxy clusters}
\ifjournal
  \titlerunning{Correlation of the magnetic field and gas density in clusters}
  \author{K. Dolag\inst{1}, S. Schindler\inst{2}, F. Govoni\inst{3,4} and 
L. Feretti\inst{4}}
  \offprints{K.~Dolag}
  \institute{Max-Planck-Institut f\"ur Astrophysik, P.O.~Box 1317, D--85741
    Garching, Germany \and
    Astrophysics Research Institute,Liverpool John Moores University,
    Twelve Quays House, Birkenhead CH41 1LD, United Kingdom \and
    Dipartimento di Astronomia, Univ. Bologna, Via Ranzani 1, I--40127 
    Bologna, Italy \and
    Istituto di Radioastronomia CNR, Via P. Gobetti,
    101 I--40129 Bologna, Italy }
  \date{}
\else
  \author{K. Dolag$^1$, S. Schindler$^2$, F. Govoni$^{3,4}$ and 
L. Feretti$^4$\\
    $^1$Max-Planck-Institut f\"ur Astrophysik, P.O.~Box 1317, D--85741
    Garching, Germany\\
    $^2$Astrophysics Research Institute,Liverpool John Moores University,
    Twelve Quays House, Egerton Wharf, Birkenhead CH41 1LD, United Kingdom\\
    $^3$Dipartimento di Astronomia, Univ. Bologna, Via Ranzani 1, I--40127 
    Bologna, Italy \\
    $^4$Istituto di Radioastronomia CNR, Via P. Gobetti,
    101 I--40129 Bologna, Italy }
  \date{}
\fi

\abstract{
We present a correlation between X-ray surface brightness and
Faraday rotation measure in galaxy clusters,  both, from radio and
X-ray observations as well as from modeling of the intra-cluster
medium. The observed correlation rules out a magnetic field of constant
strength throughout the cluster. Cosmological,
magneto-hydrodynamic simulations of galaxy clusters are used to
show that for a magnetic field of cosmic origin this
correlation is expected and excellently reproduces the
observations showing that the RMS scatter of the Faraday rotation
increases linearly with the X-ray surface brightness. 
From the correlation between the observable quantities, rotation
measure and X-ray surface brightness, we infer a relation between 
the physical quantities: magnetic field and gas density. For the best
available observations, those of A119, we find $B \propto n_e^{0.9}$. 
\ifjournal\fi
}

   \maketitle

%
%

\section{Introduction}

It is now well established that the intra-cluster  medium (ICM)
in clusters of galaxies is magnetized. The presence of
cluster magnetic fields is directly demonstrated  by
the existence  of  diffuse  cluster-wide synchrotron
radio emission (radio halos and relics)
as revealed in the Coma cluster
(Giovannini et al. 1991, 1993) and some other
clusters (e.g., Feretti 1999).
Under the assumption that the energy density within
radio sources is minimum (equipartition condition),
magnetic field values in the range 0.1-1 $\mu$G
are derived for the radio emitting regions, i.e. on scales
as large as $\sim 1$ Mpc. These values are consistent
with those suggested from the recent
detections of Inverse Compton hard X-ray emission
in clusters with halos or relics
(Bagchi et al. 1998, Fusco-Femiano et al. 1999, 2000,
Rephaeli et al. 1999).

In addition, indirect observational evidence for the existence
of cluster magnetic fields can be inferred from rotation
measure (RM) studies of extragalactic radio sources
located within or behind the clusters.
Kim et al. (1991) analyzed the RM of radio
sources in a sample of Abell clusters and found that $\mu$G level
fields are widespread in the ICM, regardless of whether there is
a strong radio halo or not.
In a recent statistical study,  Clarke et al. (2001)
found  that the ICM  in clusters is
permeated with a high filling factor by magnetic fields at levels
of 4 - 8 $\mu$G and with a correlation length of $\sim$15 kpc,
 up to $\sim$0.75 Mpc from the cluster center.
In Coma, A119 and in the 3C129 cluster, 
Feretti et al. (1995), Feretti et al. (1999) and Taylor et al. (2001)
found a magnetic field component between  5 and 10  $\mu$G, tangled
on scales of a few kpc.
In Coma, the existence of a weaker magnetic field component,
ordered on a scale of about one cluster core radius and with a strength
of 0.1-0.2 $\mu$G was also inferred.
Strong magnetic fields, up to the
extreme value of tens of $\mu$G
have been found in clusters with  ``cooling flows''
(e.g., Hydra A, Taylor \& Perley 1993; 3C295, Allen et al. 2001),
where it has been suggested that the cooling flow process
may play a role in magnetic field amplification
(Soker \& Sarazin 1990, Godon et al. 1998).

The magnetic field strengths obtained from RM arguments are
therefore higher 
than the  values derived either from the radio data,
and or from Inverse-Compton X-ray emission.
We note, however, that values deduced from radio synchrotron emission
and from inverse Compton
refer to averages over large volumes. 
Instead
RM estimates give a weighted average of the field and gas density
along the line of sight, and
could be sensitive to the presence of filamentary
structure in the cluster and/or to the existence of local turbulence around
the radio galaxies. They could therefore  be higher
than the average cluster value (see also Goldshmidt \& Rephaeli
1993). From the observational evidence, we can generally
conclude that clusters
of galaxies are pervaded by magnetic fields at least
of the order of $\sim \mu$G.
According to these findings, the
energy associated with the magnetic field is
comparable to the turbulent and thermal energy, i.e.
the fields are strong enough to be
dynamically important in a cluster.

The observations are often interpreted in terms of
the simplest possible model, i.e. in this case
 a constant field throughout the whole cluster.
However,
 Jaffe (1980) suggested that the magnetic field
distribution depends on the thermal gas  density and on the
distribution of massive galaxies and therefore would decline with
the cluster radius, as also derived by Brunetti et al. (2001) in Coma.

The knowledge of the properties of the large-scale magnetic fields in clusters
is important for studying cluster-formation and evolution, and
has significant implications for primordial
star formation (Pudritz \& Silk 1989).
The magnetic fields could be primordial (Olinto 1997),
or injected into the ICM from galactic winds or from active
galaxies (Kronberg et al. 1999, V\"olk \& Atoyan 1999).
The seed fields, whose strengths have been calculated to be up to
10$^{-9}$ G (see Kronberg 1994, Blasi et al. 1999),
are likely to be amplified by turbulence following a
cluster merger  (Tribble 1993, Dolag et al. 1999).
Amplification by turbulence excited by galactic motions
(Jaffe 1980, Ruzmaikin et al. 1989) has been shown to be
insufficient to create magnetic fields of the appropriate
strength (De Young 1992, Goldshmidt \& Rephaeli 1993).

In this paper we use the cosmological magneto-hydrodynamic code (MHD)
presented by Dolag
et al. (1999) to derive the magnetic field of clusters of galaxies.
We compare this magnetic field with the X-ray flux as a function
of distance to the cluster center.
We then perform the same comparison with quantities from observations,
i.e. the X-ray
surface brightness and the RM, for four clusters: Coma,
A119, A514 and 3C129. The correlations found, both in simulations and
in observations, allow conclusions to be drawn about the connection between
magnetic field and the gas density in clusters.

The paper is organized as follows: Sect. 2 summarizes the
simulations we have performed, Sect. 3 gives the relation between the X-ray
surface brightness
$S_x$ and the root mean square of the rotation measure
$\sigma_\mathrm{RM}$ obtained from the simulations,
Sect. 4 presents the available radio and X-ray data, which are compared
with the simulations in Sect. 5. In Sect.~6 we draw conclusions about
the correlation of the gas density and the magnetic field.

\section{Simulations}
We use 
the cosmological MHD code described in Dolag et
al.~(1999) to simulate the formation of magnetized galaxy
clusters from an initial density perturbation field.
The evolution of the magnetic field is followed starting from an
initial seed field. This field is amplified by 
compression during cluster collapse. Merger events
and shear flows, which are very common in the cosmological
environment of large-scale structure, lead to Kelvin-Helmholtz
instabilities. They further increase the field strength by a
large factor. In order to reproduce the present
$\mu$G fields, an initial field strength of $10^{-9}$G is required.
In such a scenario the final magnetic field structure is
produced by the formation of the galaxy cluster in the context of large-scale
evolution. This means that the final magnetic field
properties are predictions of our understanding of the formation
of galaxy clusters, under the assumption that the magnetic field
in galaxy clusters results from amplification of weak seed fields.
Our models are able to reproduce the observed Faraday rotation measures very
well.

\subsection{GrapeMSPH}
The code combines the merely gravitational interaction of a
dark-matter component with the magneto-hydrodynamics of a gaseous
component. 
Gravitational forces are calculated on the special-purpose hardware
component Grape 3Af (Ito et al.~1993) which is connected to the host
computer. Given a collection of
particles, their masses and positions, the Grape board computes their
mutual distances and the gravitational forces between them, smoothed
at small distances according to the Plummer law.
The gas dynamics is computed in the smoothed particle (SPH)
approximation which benefits from the list of 
 neighboring
particles also returned by the the Grape board.
It is supplemented with the magneto-hydrodynamic
equations to trace the evolution of the magnetic fields which are
frozen into the motion of the gas because of its assumed high
electric conductivity. The 
backreaction
of the magnetic field on the
gas is included.  Extensive tests of the code were successfully
performed and described in a previous paper (Dolag et al. 1999).
 $\vec\nabla \cdot \vec{B}$ is always negligible compared to
the magnetic field divided by a typical length scale of
the magnetic field within the simulations, e.g. 50-100 kpc.
The code also assumes the ICM to be an
ideal gas with an adiabatic index of 5/3 and neglects
 radiative
cooling. The surroundings of the clusters are dynamically important
because of tidal influences and the details of the merger history. In
order to account for this the cluster simulation volumes are
surrounded by a layer of boundary particles in order to
represent accurately the sources of the tidal fields in the cluster
neighborhood. The details of the code, the models and the obtained
magnetic field structure will be presented in a forthcoming
paper (Dolag et al., in preparation).

\subsection{Initial conditions}
As shown in Dolag 
(2000) 
the magnetic fields in our simulations
reproduce the Faraday rotation observations independent of the
chosen cosmology. Therefore we restrict for these comparisons
our models to one SCDM cosmology ($\Omega_{\mathrm m}^0=1.0$,
$\Omega_\Lambda^0=0$, $H_{0} = 50\,{\rm km\,s^{-1}\,Mpc^{-1}}$ and 5\%
baryon fraction).
For the cosmological initial conditions we used
a set taken from Bartelmann \& Steinmetz (1996). 
Here, we have $\sim50,000$ collisionless dark-matter particles with
mass $3.2\times10^{11}\,M_\odot$, mixed with an equal number of gas
particles whose mass is twenty times smaller. 
This central region is surrounded by $\sim20,000$ collisionless boundary
particles whose mass increases outward to mimic the tidal forces of
the 
neighboring 
large scale structure.
These ten different realizations result in clusters of different final
masses and different dynamical states at redshift $z=0$. 

They cover a temperature range between 6 and 12keV
with one very big cluster even reaching 20keV.
To gain a larger range in temperatures of our simulated cluster
sample, we also use ten less massive objects identified close to the main
clusters. They extend the temperature range down to 2keV.
As they are more poorly resolved in the
simulations, the results from the smaller objects have to be treated
with some care.

Each of these clusters was simulated using five
different models for the initial magnetic field setup as described
below, leading to a set of 50 simulations.

\begin{table}[t]
\begin{center}
\begin{tabular}{|c|c|c|c|c|c|}
\hline
 model & $\langle B_\mathrm{final}\rangle_\mathrm{core}^\mathrm{SCDM}$ &
 $\alpha_{10}$ & $\alpha_{20}$ & $\delta$ & RMS\\
\hline
low     & $0.4\;\mu{\rm G}$ & 0.93 & 0.88 & 2.0 & 2.7 \\
chaotic & $0.4\;\mu{\rm G}$ & 0.92 & 0.87 & 2.3 & 2.3 \\
double  & $1.0\;\mu{\rm G}$ & 0.94 & 0.90 & 2.8 & 2.5 \\
medium  & $1.1\;\mu{\rm G}$ & 0.93 & 0.87 & 1.6 & 2.3 \\
high    & $2.5\;\mu{\rm G}$ & 0.91 & 0.87 & 1.1 & 1.7 \\
\hline
\end{tabular}
\end{center}
\caption{Final magnetic fields (column 2) for the different
models in the SCDM cosmology.
The values are averages over the ten main clusters for each model,
where the mean field within ten percent of the virial radius
is measured. Column 3 and 4 give the mean value of the calculated
slope of the correlation over the ten main clusters and including
the ten smaller objects, respectively, as explained in the text. Column 5
shows the scaling $\delta$  
of the normalization with cluster temperature. The
last column gives the RMS of the individual 
clusters around this scaling as seen as scatter of the data
points around the best fit for the medium model in the right panel of 
Fig.~\ref{fig:lx_parameter}.}
\label{tab:bfield}
\end{table}

In the absence of any detailed knowledge on the origin of primordial
magnetic seed fields, we explore a whole set of initial magnetic
field configurations. To study the effect of changing the mean energy
density of the magnetic seed field we varied this in our
``{\em homogeneous}''
simulations labeled as ``{\em low}'', ``{\em medium}'' and ``{\em
high}'', taken 0.2,1 and $5\times10^{-9}$G as initial field
at $z=15$ respectively.
Current observations (Clarke et al. 2001), compared with synthetic
Faraday rotation measurements from our simulations, fall between the
``{\em medium}'' and ``{\em high}'' sets when comparing the
amplitude and the radial shape of the rotation measure signal
produced by clusters.
 We also compare two extreme
models for the initial field configuration.
In one case (``{\em homogeneous}''), we
assume that the field is initially constant throughout the simulation
volume. In the other case (``{\em chaotic}''), we let the initial
field orientation vary randomly  on our initial grid, subject only
to the condition that $\nabla\cdot\vec B=0$.
As a test we
performed also a set of simulations in which we doubled
the mass resolution, labeled as ``{\em double}''. The resulting magnetic
field within our clusters is summarized in Table~\ref{tab:bfield}.

By projecting along the spatial directions, every cluster
gives us three independent maps. Therefore, including the ten smaller
objects, we gain a set of sixty maps for each initial magnetic field
configuration.

\section{The synthetic $\sigma_\mathrm{RM}$-S$_{\bf x}$ correlation}

\begin{figure}[t]
 \includegraphics[width=0.49\textwidth]{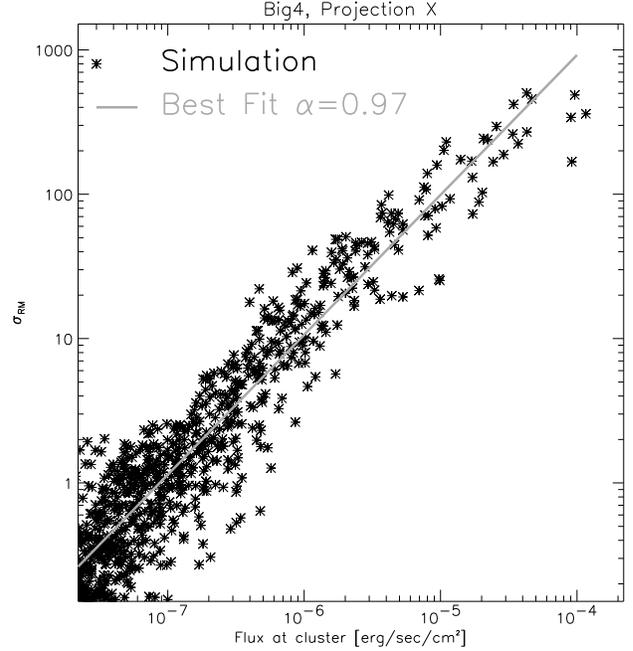}
\caption{The symbols show a point by point comparison of the X-ray surface
brightness and the RMS of the synthetic rotation measure calculated
from one projection of one simulated cluster taken from the
``{\em medium}'' models. Overlayed is the best
fit power-law. Clear to see that the correlation follows closely a
power-law with a slope around unity.} \label{fig:lx_synth}
\end{figure}

Our models predict the structure of the
magnetic field in galaxy clusters within the assumed scenario. One
prediction is, that the magnetic field decreases with increasing distance
from the center. Therefore we
can predict the statistics of the rotation measure within our models.
Comparing this with the X-ray emission provides an opportunity to
measure the distribution of the magnetic field in galaxy clusters.
Two 
quantities are  obtained from the models at different positions in the
cluster:  the X-ray flux calculated as line of sight integral over
the emissivity within the energy range
$[0.1,2.4]$keV, which translates to an X-ray surface
brightness, and the root mean square (RMS) 
of the synthetic rotation measure (RM). To this aim,
 the synthetic maps are divided in 125kpc x 125kpc boxes,
in which the RMS of the rotation measure and the mean of the X-ray flux
are calculated (see Fig.~\ref{fig:lx_synth}).
Here, we compare the two line of sight integrals
\begin{equation}
   S_x \propto
   \int n_{\mathrm e}^2 \; \sqrt{T} \; {\mathrm d} x
   \Longleftrightarrow
   \sigma_\mathrm{RM} \propto
   \int n_{\mathrm e} \; B_\| \; {\mathrm d} x
\end{equation}

with each other.  In comparing the two quantities, we obtain the
correlation of magnetic field versus density, when neglecting
the temperature dependence on the left side.
Actually, the dependence of the X-ray emission on
the temperature is even flatter than the square root between 1 and
10 keV due to line emission. For the ROSAT energy range (0.1-2.4 keV)
the temperature dependence can be neglected.
We find a clear correlation between
rotation measure and X-ray flux
in all our simulated clusters suggesting that the magnetic field is
directly related
to the gas density.
Fitting this correlation by

\begin{equation}
\sigma_\mathrm{RM}=A\left(
   \frac{S_x}{10^{-5}\mathrm{erg}/\mathrm{cm}^2/\mathrm{s}}\right)^\alpha,
\end{equation}
the slope $\alpha$ somewhat
depends on the cutoff in surface brightness and rotation measure,
applied to the synthetic data. The cutoff was chosen to
be $10^{-4}$ of the maximum
found in the individual maps. The fits tend to give values slightly
smaller than one, but all the individual correlations are still
consistent with a slope of one.

\begin{figure*}[t]
 \includegraphics[width=0.99\textwidth]{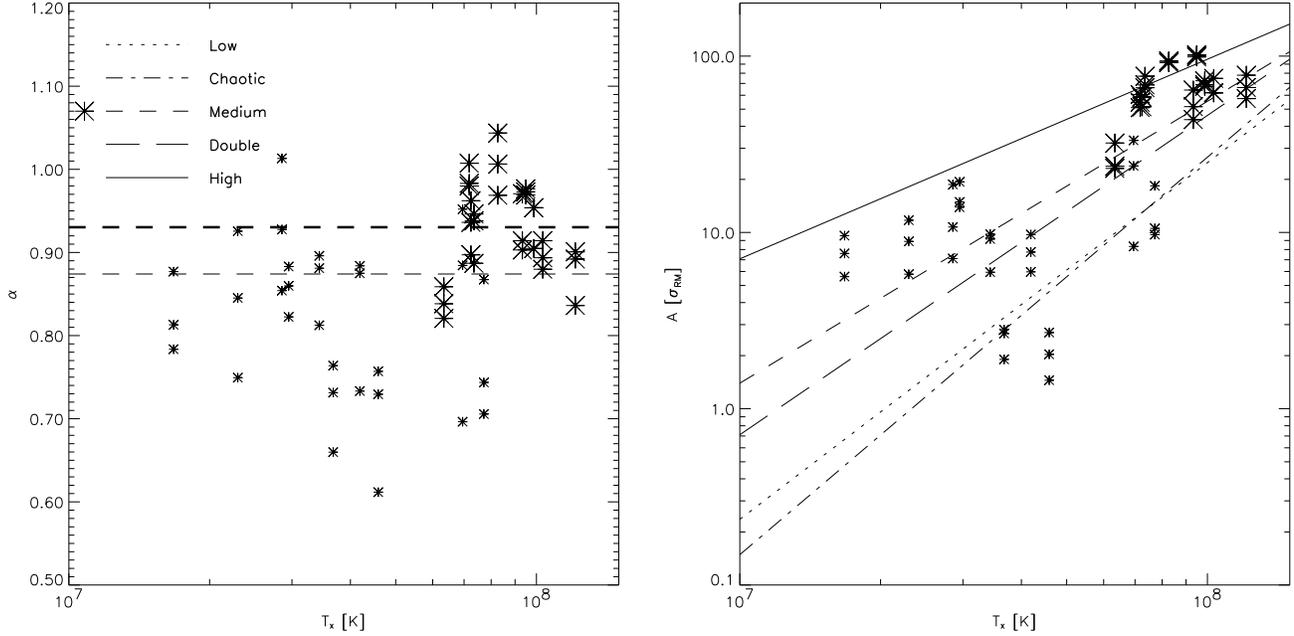}
\caption{In the left panel the fitted slope $\alpha$ of the
$\sigma_\mathrm{RM}$-$S_x$-correlation as
function of the cluster temperature is plotted. Here fits in all three
projection directions of all twenty model clusters are shown. The symbols are
plotted only for the ``{\em medium}'' models.
The large symbols are the ten main clusters, the smaller symbols are the smaller
objects mentioned before. The lines mark the average over the ten main
clusters (bold line) and the average over all 20 objects (thin line). 
The right side shows the normalization 
$A$
of the correlation function
of these power-law fits for the same models. The different lines are
for different initial field configurations as labeled in the plot.
} \label{fig:lx_parameter}
\end{figure*}

Figure \ref{fig:lx_parameter} shows the results of the slope $\alpha$
for various cluster models and for different initial magnetic field
configurations in our simulation. To increase the temperature range
of clusters used for this analysis we also include the ten smaller
objects found near the main
clusters in our simulations as described before. In total we use
twenty different objects for every initial magnetic field
configuration, in which we 
analyze 
three spatial projection directions.
The left panel of Fig.~\ref{fig:lx_parameter} shows that the slope
of the correlation is constant over the temperature range of our
simulations. The symbols are taken from the ``{\em medium}'' models,
the different lines are the best fits to different magnetic field
configurations as labeled in the plot. It is obvious that the slope
did not change for different initial field strengths and orientations.
The large symbols are the ten main clusters, the smaller symbols are
the smaller objects.
Also the clusters with doubled mass resolution show the same slope.
There is a small trend visible, that when including the smaller
objects, the slope tends to be somewhat smaller. This is only 5\% and
may well be due to the poor resolution in these smaller objects and
due to the chosen cutoff for the fitting procedure. Also here,
a slope of 1, yields a good fit to data. The obtained values for
the fits to the synthetic data are summarized in Table~\ref{tab:bfield}.

The simulations not only predict that the magnetic field
scales similarly to the density within all clusters but also
show that clusters have different central magnetic field strengths
depending on their dynamical state and their temperature, leading to
an offset of this correlation.
Therefore the normalization 
$A$
of the
$\sigma_\mathrm{RM}$-$S_x$-correlation  in Eq. 2
should depend on
the ``global'' magnetic field strengths (determined in the simulations
by choosing the initial field strengths), the temperature of the
cluster and its dynamical state.
Analyzing the whole set of clusters, we expect to see a trend
of the normalization with cluster temperature, whereas the scatter
around this trend should be due to the individual
dynamical states of the clusters.

The right panel of Fig.~\ref{fig:lx_parameter} shows, as expected,
that the normalization $A$ of this correlation rises with temperature and
increasing ``global'' magnetic field strengths, but does not change for
different field configurations or when doubling the mass resolution.
The symbols are again taken from the ``{\em medium}'' models, the
different lines are the best fit of
\begin{equation}
   A\propto T^\delta
\end{equation}
for all the different models as labeled in Fig.~\ref{fig:lx_parameter}.
The values obtained for these fits are also summarized in
Table~\ref{tab:bfield}. The large scatter of the individual objects around
this trend also suggests, that the dynamical state of the individual
objects plays a crucial role. This reflects the established fact
that merging of clusters strongly amplifies the magnetic field
(Roettiger et al 1999).
Table~\ref{tab:bfield} (column 6) also summarize the calculated RMS 
of the individual clusters around this scaling of the normalization $A$.

\section{Data Presentation}

\begin{table*}
\caption{Summary of cluster data used for the analysis 
}
\begin{center}
\begin{flushleft}
\begin{tabular}{cllllll}
\hline
\noalign{\smallskip}
Cluster &z & Redshift  & $n_\mathrm{H}$ & Temperature & Temperature  & Temperature\\
        &  & reference & $[10^{21}]$    & [keV]       & reference    & used\\
\noalign{\smallskip}
\hline
\noalign{\smallskip}
A119 &0.0443&Struble \& Rood (1999)&0.289& $5.1^{+1.0}_{-0.8}$ & Edge et al. (1990) &\\
     &      &                      &     & $5.9^{+1.1}_{-0.9}$    & David et al. (1992) & \\
     &      &                      &     & $5.6$                  & Markevitch et al. (1998) & X \\
A514 &0.0714&Fadda et al. (1996)   &0.321& $3.6^*$                & Arnaud \& Evrard (1999) &\\
Coma &0.0232&&0.0895& $8.2$                  & Arnaud et al. (2001) & X\\
3C129&0.021 &&7.1  & $5.5\pm0.2$            & Leahy \& Yin (2000)   &  \\
     &      &                      &     & $5.6^{+0.7}_{-0.6}$    & Edge \& Steward (1991) & X  \\
     &      &                      &     & $6.25^{+0.27}_{-0.26}$ & Taylor et al. (2001) &  \\
\noalign{\smallskip}
\hline
\label{tab_cl}
\end{tabular}
\end{flushleft}
Caption:
Column 1: Cluster name;
Column 2: Redshift;
Column 3: Redshift reference;
Column 4: $n_\mathrm{H}$ from Dickey \& Lockman (1990);
Column 5: Temperature;
Column 6: Temperature reference;
Column 7: Temperature used for the analysis;
$^*$ derived from $L_x-T$ relation.
\end{center}
\end{table*}

To analyze in detail the magnetic field structure
in clusters of galaxies it is crucial to compare
the correlation between the X-ray emission
and the synthetic $\sigma_\mathrm{RM}$ with the
relation obtained from the data.

For such a comparison we need highly polarized radio galaxies,
located at different
distances from the cluster center, for which the rotation measure and the
contribution of the inter-galactic medium to the X-ray emissivity
at the source position
are known.
The cluster A119 is the most important one for our analysis, as it
contain three sources spanning a useful range of line of sights through
the cluster. Additionally sources in A514, Coma and 3C129 are used
for a combined analysis.
Following, we give a brief description of the X-ray and radio data
used here.
While the radio data were taken from previous work,
we re-analyzed X-ray archival data to measure the exact surface
brightness at the position of the radio galaxies.

\subsection{X-ray data}

For the four clusters X-ray data obtained with ROSAT were retrieved
from the archive.
A 15 ksec ROSAT/PSPC observation was used for A119 
in the hard band (0.5-2.0~keV).
For A514 an 18 ksec ROSAT/PSPC observation 
(hard band) was used.
The pointing contains many point sources which are probably not
associated with the cluster. These point sources were removed
before smoothing by replacing the pixel values with values taken
from the surrounding area.
For the Coma cluster we used a 21 ksec ROSAT/PSPC observation. Also
here only the hard band data were taken into account.
For 3C129 two ROSAT/HRI 
pointings were
used with a total
exposure time of 39 ksec.
For all clusters the X-ray values were extracted from
images with a Gaussian smoothing of $\sigma$=40$''$.
The backgrounds were taken from empty regions in the pointings. For
the cluster 3C129 the background is higher, because it is an HRI
observation. This high background introduces a large uncertainty into
the countrate of 3C129 of about 50\%.
For the other clusters the 
errors in the countrate
are statistical errors inferred from the numbers of photons observed
at the position of the radio sources. For the source in the Coma
cluster additionally note that the X-ray flux over the region of the
radio source varies by $\pm5$\%. This variation is taken into account
in the error listed in Table \ref{tab1}.

A summary of the clusters we used can be found in Table~\ref{tab_cl}.
Note that for some clusters different temperatures are given in
literature. As no good temperature measurement exists for A514 the
temperature was estimated from the
$L_x-T$ relation (Arnaud \& Evrard 1999).

\subsection{Radio data}
\begin{table*}
\caption{Summary of RM and X-Ray data}
\begin{center}
\begin{tabular}{cccccccc}
\hline
\noalign{\smallskip}
Cluster &Radio source & Dist    & $<$RM$>$     & $\sigma_\mathrm{RM}$ & X-ray
        (PSPC) & Background & Flux\\
        &             & \arcmin & rad m$^{-2}$ & rad m$^{-2}$  & $10^{-3}$(cts/sec/arcmin$^2$)
                      & $10^{-3}$(cts/sec/arcmin$^2$) & $10^{-5}$(erg/sec/cm$^2$)  \\
\noalign{\smallskip}
\hline
\noalign{\smallskip}
A119      & 0053-015       &2.4 &+28 &$152^{+30}_{-42}$&$ 7.3\pm0.6 $&0.27&$2.4\pm0.19$\\
$''$      & 0053-016       &6.4 &-79 &$91 ^{+20}_{-33}$&$ 4.0\pm0.5 $&$''$&$1.3\pm0.14$\\
$''$      & 3C29           &21.4&+4  &$13 ^{+14}_{-13}$&$ 0.8\pm0.2 $&$''$&$0.26\pm0.07$\\
A514      &J0448-2025      &2.7 &+104&$63 ^{+16}_{-41}$&$2.43\pm0.31$&0.19&$0.86\pm0.12$\\
$''$      &J0448-2032      &7   &+46 &$47 ^{+11}_{-21}$&$2.09\pm0.30$&$''$&$0.76\pm0.10$\\
$''$      &J0448-2032-N$^*$& 6.6 &+25 &$54 ^{+12}_{-21}$&$2.26\pm0.30$&$''$&$0.82\pm0.11$\\
$''$      &J0448-2032-S$^*$& 7.3 &+56 &$38 ^{+10}_{-23}$&$2.04\pm0.29$&$''$&$0.74\pm0.10$\\
3C129$^\P$&3C129.1         &0   &+21 &$200^{+39}_{-50}$&$2.25\pm0.32$&3.5 &$7.0\pm1.0$\\
$''$      &3C129           &16  &-125&$82 ^{+17}_{-25}$&$0.73\pm0.28$&$''$&$2.3\pm0.9$\\
Coma      &NGC4869         &5   &-243&$87 ^{+36}_{-48}$&$25.4\pm2.7 $&0.69&$5.95\pm0.6$\\
\noalign{\smallskip}
\hline
\label{tab1}
\end{tabular}
\end{center}
Caption.
Column 1: Cluster name;
Column 2: Source name;
Column 3: Distance from the cluster center;
Column 4: Average value of RM;
Column 5: RM dispersion;
Column 6: X-ray surface brightness in the ROSAT hard band,
background corrected;
Column 7: X-ray background used;
Column 8: X-ray flux at the position of the radio source;
$^*$Source divided into North and South lobe;
$^\P$ HRI counts;
\end{table*}

The rotation measures of the radio sources in 
A119, A514, Coma and 3C129 have been obtained
with the  Very Large Array (VLA), using sensitive
data at multiple wavelengths.

Linear polarized electromagnetic radiation passing
through a magnetized ionized medium suffers a rotation of
the plane of polarization:
\begin{equation}
\Psi(\lambda) =\Psi_0 + \lambda ^2 RM
\end{equation}
where  $\Psi(\lambda)$ is the
position angle observed at a wavelength $\lambda$ and the $\Psi_0$
is the intrinsic position angle.
The position angle of the plane of polarization is an observable
quantity, therefore,
images of rotation measure can be constructed,
by linear fitting the polarization angle as a function of $\lambda ^2$.

In the cluster A119, the polarization properties of 
three extended radio galaxies were analyzed (Feretti et al. 1999).
The two sources 0053-015 and 0053-016 show a head-tail
structure of about 5\arcmin~in size, and are projected close to the
cluster center, The third source, 3C29 (0055-016), is a typical FRI,
about 2.5\arcmin~ in size and located at the cluster periphery.
In the cluster A514, two extended radio sources are suitable
for a polarization study (Govoni et al., in preparation), owing to 
their high degree of polarization: 
 J0448-2025 is a head-tail
radio source of about 0.8\arcmin~ in size while J0448-2032
is an FRI radio galaxies with an projected extension of about
1.4\arcmin.
In the Coma cluster, the tailed radio galaxy NGC4869 
was analyzed by Feretti et al. (1995).
It is located near the cluster center and it is 
extended about 4\arcmin.
In the 3C129 cluster, two radio sources were analyzed
by Taylor et al. (2001): 3C129.1 at the cluster center, and
the tailed radio galaxy 3C129 at the cluster periphery.

To estimate the errors on the dispersion of the rotation measure, 
we assumed the $\sigma_\mathrm{RM}$ composed as follows:
\begin{eqnarray}
    (\sigma_\mathrm{RM})^2 &=& 
      (\sigma_\mathrm{RM}^\mathrm{measured})^2
     -(1\pm\eta_1)(\sigma_\mathrm{RM}^\mathrm{noise})^2
\nonumber\\    
&\pm&2\sigma_\mathrm{RM}^\mathrm{measured}\sigma_\mathrm{RM}^\mathrm{noise}/\sqrt{N}\nonumber\\
     &\pm&\eta_2(\sigma_\mathrm{RM}^\mathrm{measured})^2.
\end{eqnarray}
Here, the first term $\sigma_\mathrm{RM}^\mathrm{measured}$ is the
signal extracted from the map. The second term describes the widening of the
signal due to the uncertainties
$\sigma_\mathrm{RM}^\mathrm{noise}$ within the 
individual pixels of the maps, where $\eta_1$ reflects our lack of
knowledge of this value. We 
chose 
$\sigma_\mathrm{RM}^\mathrm{noise}$ between
20 and 30 for the individual maps and a value of 0.5 for $\eta_1$, as
these errors are not known very precisely. The third term reflects
the statistical error inferring
$\sigma_\mathrm{RM}^\mathrm{measured}$
from the given distribution of $N$ independent resolution
elements across the source and is
small due to the fact that $N$ is moderately high. The last term reflects the
uncertainty of the exact source position along the line of sight,
characterized by the value of K in Eq. (9). We chose
$\eta_2=(624-411)/411=0.52$ as the typical value. Note that in
principle the fourth
term only contributes when the rotation measure is deprojected.

For each radio source, Table~\ref{tab1} shows the distance
from the cluster center, the average value of the RM, the
$\sigma_\mathrm{RM}$, the X-ray surface brightness,
X-ray background and the X-ray flux
of the cluster calculated in the position of the radio source.
In both clusters, A119 and A514,
the value of the $\sigma_\mathrm{RM}$
decreases with increasing projected distance from the cluster center.
These data are 
good evidence for the existence of a magnetic field
associated with the intra-cluster medium which 
contributes
to
the Faraday rotation according to how much magneto-ionized medium is
crossed by the polarized emission.

\subsection{Modeling the magnetic field}

Various physical models have been invoked to explain the rotation
measures observed in radio sources belonging to clusters of
galaxies (Lawler \& Dennison 1982, Tribble 1991, Felten 1996).

Assuming a magnetic field topology with cells of constant size,
density and magnetic field strength, but
random orientation inside each cell,
the contribution to the RM from each cell is given by:
\begin{equation}
RM=812n_eB_{\parallel}l~~~(rad~m^{-2})
\end{equation}
where $n_e$ is the electron density in cm$^{-3}$, $B_{\parallel}$ is the
magnetic field component along the line of sight measured in $\mu$G, and
$l$
is the cell size in kpc.
The observed RM, along any given
line of sight, will be generated by a random walk process.
Thus, the distribution of RM can be modeled with a Gaussian
with 
mean
value of 0 ($rad~m^{-2}$) and dispersion given by:
\begin{equation}
 \sigma_\mathrm{RM}=\frac{812}{\sqrt{3}}n_eBN^{\frac{1}{2}}l
\end{equation}
where N is the number of cells along the line of sight.

Considering a similar model but with a density distribution
that follows a $\beta$-profile:
\begin{equation}
n_e(r) =n_0 (1 + r^2/r^2_{\rm c})^{-3 \beta/2}
\label{betam}
\end{equation}
where $n_0$ is the central gas density, $r_{\rm c}$ is the
core radius, and  $\beta$ is the slope of the profile,
Felten (1996) derived the following
relation for the RM dispersion:

\begin{equation}
\sigma_\mathrm{RM}= {{K B n_0  r_c^{1/2} l^{1/2} }\over
{(1+r^2/r_c^2)^{(6\beta -1)/4}}} \sqrt {{\Gamma(3\beta-0.5)}\over{\Gamma
(3\beta)}}
\label{srm}
\end{equation}
where $\Gamma$ is the Gamma function (while
the other parameters have the same meaning 
as
in the previous
formula).
The factor $K$ depends on the integration path over
the gas density distribution:
$K$ = 624, if the source lies completely
beyond the cluster, and $K$ = 441 if the source lies
at the cluster center.

\begin{figure*}[t]
 \includegraphics[width=0.49\textwidth]{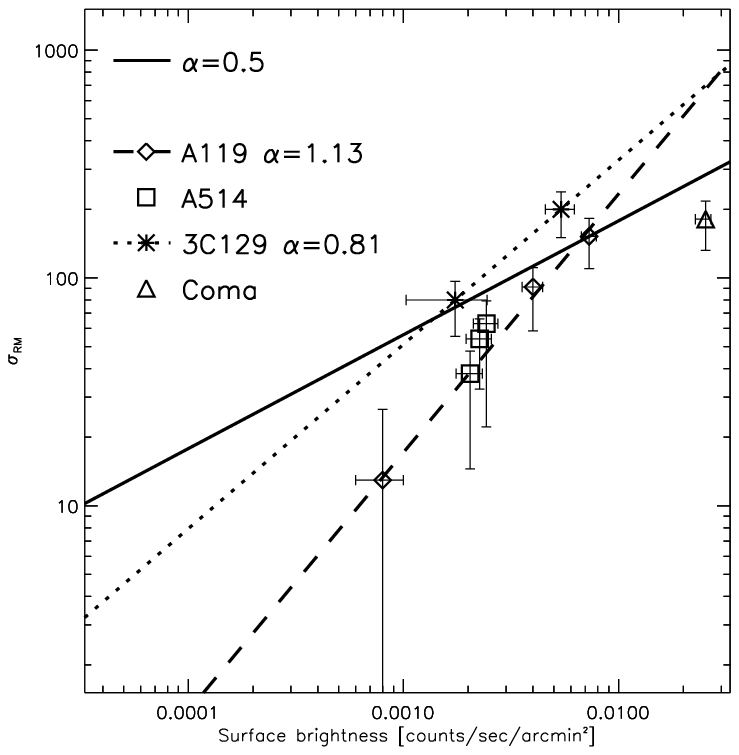}
 \includegraphics[width=0.49\textwidth]{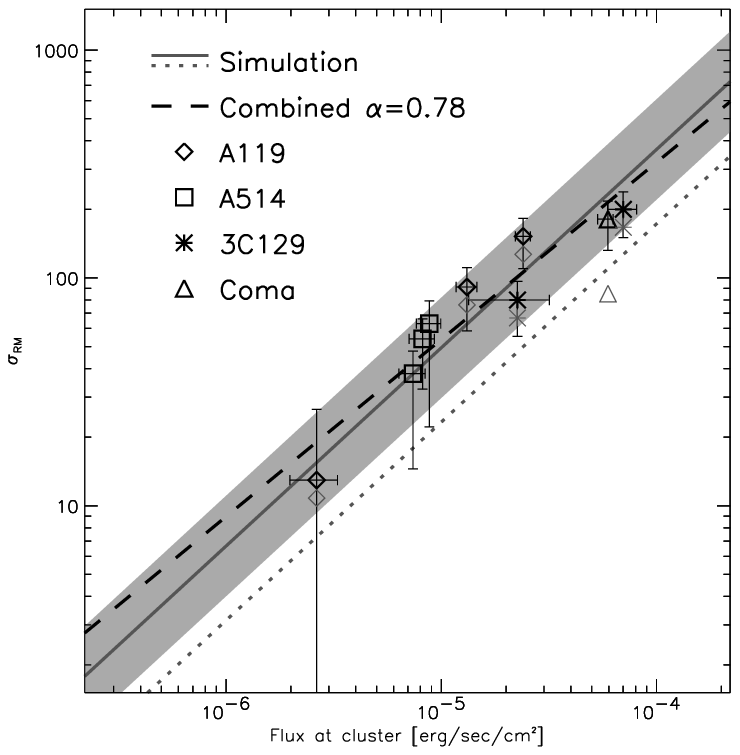}
\caption{Shown is the measured correlation between the X-ray surface
 brightness and the RMS of the rotation measurement (left panel).
The symbols are the measurements from the clusters as indicated in the
plot. For 3C129 the HRI counts are converted to PSPC hard band counts
with a factor of 2.4.
The dashed and the dotted line show the best fit to the
individual clusters A119 and 3C129.
The solid line shows the expected correlation for an overall constant
magnetic field inside the cluster adapted to the data from A119.
The right panel combines
all measurements by converting the surface brightness to the position
of the cluster. The expected correlation for a 5keV cluster from the
simulations is shown for the ``{\em high}'' models (solid line) and
for the ``{\em medium}'' model (dotted line). The gray region marks
the scatter for individual clusters expected for the simulations of
the ``{\em high}'' models. The combined fit to the data is shown as 
dashed line. The gray symbols indicate, where the data points move 
when we correct for the cluster temperature according to the
theoretical models.
} \label{fig:lx_true}
\end{figure*}

By combining  Eq. (\ref{srm})
with the X-ray surface brightness profile
obtained from a $\beta$ model:
\begin{equation}
S(r)=S_0(1+r^2/r_c^2)^{-3\beta+0.5}
\end{equation}
and assuming that the factor $K$ is the same for all sources
in a cluster,
we expect a power-law correlation between the 
$\sigma_\mathrm{RM}$ and the X-ray brightness
with an index $\alpha$=0.5. This slope of the
correlation is independent of the values fitted for $\beta$
and $r_{\rm c}$. This expected relation, obtained assuming a
constant magnetic field and cell size is plotted in Fig.~\ref{fig:lx_true}
as a solid line.

\section{Comparing observations}

In order to test whether the correlation found in model clusters is
also present in real clusters we use observational radio and X-ray data.
In principle,
each cluster with enough radio source
polarization measurements would allow us to
determine the correlation between $S_x$ and
$\sigma_\mathrm{RM}$ individually and therefore give
independent estimates, how the magnetic field scales with density
in these objects. As there are only a small number of observations,
even spread over several clusters, a combined analysis of these
clusters is necessary. Once there are enough measurements
their combined analysis could tell, how the magnetic field
scales with the cluster temperature.

\subsection{The observed $\sigma_\mathrm{RM}$-S$_x$ correlation}
The values for the individual sources within the
clusters like positions, X-ray counts and RM measurements can be found
in
Table~\ref{tab1}. In the left panel of Fig.~\ref{fig:lx_true} we show all
data points drawn from our clusters. Plotted
is $\sigma_\mathrm{RM}$ versus the X-ray surface brightness.
Here we can try
to fit the correlation in the individual clusters as indicated in the
plot. The surface brightness is background corrected. The absorption
plays no role for the slope calculated for the individual clusters
as the counts for all sources within one cluster would have to be
multiplied by the same value, leading only to a shift of this
correlation.
A119 (diamonds) has the best individual data, and the best fit
relation gives a slope of $\alpha=1.13$ which is comparable to what we
expect from the simulations. 
In 3C129 the lower data point has a large uncertainty in the X-ray
measurement, as it is noise dominated. 
The correlation referred from these two 
data points
gives a slope of 0.81, which is 
to
within the uncertainties also
consistent with the expectation from the simulations.
The three data points for A514 are too close
to 
infer seriously
a slope, but they also suggest a slope
larger than one i.e. far from the $\alpha=0.5$ slope for the simple models.
For Coma (triangle) only one data point is available, therefore
it can only be used in the combined analysis.
For comparison a line with the slope of 0.5 is added as solid
line. The amplitude is chosen 
such
that it matches the innermost 
data point of A119, which has the highest values and therefore is the
most reliable data point.
Due to the lack of data, only A119 and marginally 3C129
allows us to determine the correlation from individual clusters,
but both show consistently that the slope in real clusters is close to
the expected slope from the simulations, rather than the expected
slope of oversimplified models.

\subsection{Fitting the combined Data}

For combining all measurements we have to convert the X-ray measurements to
quantities at the clusters 
themselves
to be independent of distance and
absorption. To be able to compare them with the simulations, we chose
to convert the surface brightness to the position of the cluster taking into account
the distance, the cluster temperature and the absorption by Galactic hydrogen
$n_{\rm H}$.
The values we used for the clusters can be found in
Table~\ref{tab_cl}, where column 7 marks which of the different
temperatures for individual clusters we have chosen from
literature. The calculated flux for the individual sources 
can
be found in Table~\ref{tab1}. The results are shown in the right panel of
Fig.~\ref{fig:lx_true} (thick symbols).

The gray solid line represents the expected correlation for a simulated
5keV cluster taken from the ``{\em high}'' models. The gray region
marks the expected RMS for individual clusters around this prediction,
as given in the last column of Table~\ref{tab:bfield}.
The dotted line is the expected correlation for 
a 
5keV cluster taken
from the ``{\em medium}'' models. Taking the shift of the correlation
with cluster temperature within the ``{\em medium}'' model, the solid 
line would represent 
an
8keV cluster.

All measurements combined follow very well the correlation expected from the
simulations. The combined fit gives a marginally
smaller slope of 0.78 (dashed line), but the data are still
incompatible with a slope of 0.5.

Suggested by the simulations, we have to take 
into account
a possible scaling of
the magnetic field with cluster temperature
when
we want to combine the measurements from individual clusters.
In the case of enough
measurements, the combined data could be used to infer how the magnetic
field in different clusters scales with their temperature by looking
for the scaling which leads to the smallest scatter in the
combined correlation.  With the limited data currently available,
we can only sketch how to apply the calibration of this
correlation, according to the theoretical models. As 
previously found (see Fig.~\ref{fig:lx_parameter}),
the normalization of this correlation scales as
\begin{equation}
 A  \propto \left(\frac{T}{5\mathrm{keV}}\right)^{\delta}.
\end{equation}
As the value of $\delta$ changes with the magnetic field model used
in the simulations, we took a $\delta=1.6$ from the ``{\em medium}'' model,
but the difference when taking the slightly different slopes from
the other models is 
marginal.
As all clusters have measured temperatures
larger than 5keV
this calibration shifts the data points towards lower rotation measure
shown as thin, gray symbols in Fig.~\ref{fig:lx_true}. For A514 there is no direct
measurement for the temperature, only an estimate for the
temperature from the $L_x-T$-relation. Therefore we did not apply the
correction to A514. 

As the source within Coma is the only data point, which
is significantly changed when applying the theoretical
correction, it is not possible to 
test really 
if the
magnetic field changes with cluster temperature.
When correcting for temperature according to the simulations,
the RMS measurement for Coma drops somewhat below the expected
correlation. As this is only one data point, and we expect a
large scatter from the 
simulations. This
lies within the
expectations, and no definite conclusion could be drawn
from the data available at the moment.

The combined measurements follow the theoretical
expectation from our simulations very well. This holds for the individual
clusters as well as for the combined measurements. If we take
the small number of available data into account, the differences
between the fits for individual clusters themselves, the combined
measurements and the simulations are very small. They also 
indicate, that the slope of unity inferred from the simulations
is clearly favored by the
data with respect to a slope of 0.5 inferred from simple models.
The comparison of the observed $\sigma_\mathrm{RM}$-$S_x$ 
correlation with the simulations shows, that
a magnetic field somewhere between the ``{\em medium}'' and ``{\em
high}''  models 
reproduces
the observations best. 

\subsection{Implication on measuring the magnetic field B}

We can extend the 
modeling 
of the magnetic field in Sect 4.3 to the
case of variable magnetic field within the cluster.
Assuming, that the magnetic field
$B$ scales with the density $n_e$ as 
\begin{equation}
B \propto (n_e)^\gamma, 
\end{equation}
$\beta$ in Eq. (\ref{srm}) must be replaced by
$\beta(1+\gamma)$. Combining this modified
Eq. (\ref{srm}) with (10)
yields the slope of the $\sigma_\mathrm{RM}$-S$_{x}$ 
correlation to be
\begin{equation}
   \alpha=0.5\frac{3\beta(1+\gamma)-0.5}{3\beta-0.5}
\label{alpha}
\end{equation}
with 
$\gamma$ being the slope of the B-n$_e$ relation.
For a constant magnetic field ($\gamma=0$) this yields
an index of $\alpha$=0.5, 
 independent of 
the values of $\beta$ and $r_{\rm c}$, 
as already shown in Sect. 4.3.
The lines in Fig.~4 allow for a direct inference of the B-n$_e$ slope
$\gamma$ for measured $\alpha$ and $\beta$.
Assuming $\beta =0.7$ one finds  a
$\gamma$ between 0.5 and 0.9 for the observed slopes
$\alpha$ between 0.8 and 1.1. 
This is still a simplified model, as 
a constant scale length is assumed. 

\begin{figure}[t]
 \includegraphics[width=0.49\textwidth]{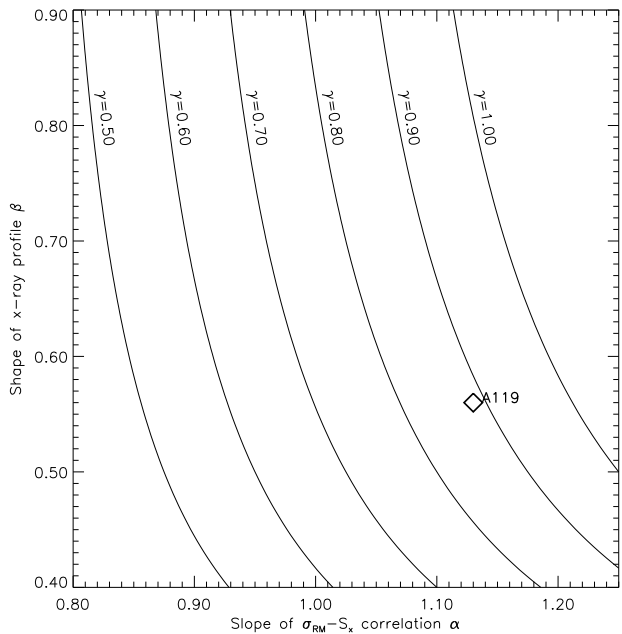}
\caption{Conversion of the slope of the $\sigma_\mathrm{RM}$-S$_{x}$ 
correlation $\alpha$ to the slope of the B-n$_e$ relation $\gamma$ for
given values of the X-ray profile shape parameter $\beta$. The
parameters of A119, $\alpha$=1.13 and $\beta$=0.56 (shown as diamond),
a value for $\gamma$ of 0.9 is inferred.
}
\end{figure}

For A119
a $\beta$=0.56 is obtained from the X-ray data
and a slope $\alpha$=1.13 is derived from the fit to the
$\sigma_\mathrm{RM}$-S$_x$ correlation
(see Fig. 3). These values yields a $\gamma= 0.9$.
For a cell size of
10-20kpc Feretti et al. (1999) calculated a magnetic field $\approx
5\mu$G using the constant 
magnetic field approximation. Taking into account the model allowing
for a varying magnetic field 
we obtain a 
central magnetic field of $\approx 7.5 \mu$G, which
decreases with radius according to $(n_e)^{0.9}$.

For 3C129 a $\beta=0.47$ is measured (Taylor et al. 2001) and we infer
a slope $\alpha=0.81$ from drawing a line through the two points
of the $\sigma_\mathrm{RM}$-S$_x$ correlation (see Fig. 3). 
This results in a $\gamma=0.5$, but -- due to the poor X-ray data of the 
lower data point -- this value has a large uncertainty.

\section{Conclusions}

We find a correlation between the magnetic field and the intracluster gas
density in galaxy clusters both, in our MHD simulations and in
observational data taken from 
literature. The results rule out
a magnetic field that is constant 
within a 
cluster. Instead we
find in the simulations a correlation between the RMS of rotation measure 
and the X-ray emission in galaxy clusters with a slope of 1.
Using all observational data currently available, 
we 
show 
that this correlation
is indeed observed in individual clusters, in particular in A119, 
where the observations show a slope close to unity.

The combination
of the measurements of 4 clusters suggests that
this correlation is likely to be universal. The
analysis of the combined data
leads to a slightly lower slope of this correlation ($\alpha$=0.8), which
may be due to the expected scatter between individual clusters, but
is still compatible with the simulations and certainly excludes
an overall constant magnetic field. 

The observational quantities, rotation measure and X-ray flux, can be
transformed to the physical quantities, magnetic field and
intra-cluster gas density, with the help of the $\beta$-model.
The relation between the slopes (Eq.~13) yields $B \propto
n_e^{\gamma}$ with $\gamma=0.9$ for A119. For 3C129 we infer a $\gamma
= 0.5$ with large uncertainty. 

Unfortunately, with the limited 
measurements presently available,
it is not possible to infer
if and how the magnetic field in different clusters depends on
other cluster properties like temperature.

\begin{acknowledgements}
We thank Torsten En\ss lin for stimulating discussions
and the referee Greg Taylor for a very useful and fast report. 

\end{acknowledgements}
%
%

\end{document}